\documentclass{llncs}

\usepackage{amsmath}
\usepackage{url}
\usepackage{listings}
\lstdefinelanguage{babellanguage} 
{morekeywords={lazy, concurrent, match, case, exception, begin, end, random, await, true, false, infinity, val, def, this, object,if,then,else,elseif,memoize,for,in,yield,while,do,with,nil,fn,let,var,datatype,type,of,and,fun,raise},
sensitive=true, 
morecomment=[l]{//}, 
morecomment=[s]{/*}{*/}, 
morestring=[b]", 
} 

\lstnewenvironment{babellisting} 
{\small} 
{} 

\newcommand{\bsrc}[1] {\lstinline!#1!}

\begin{document}

\title{Purely Functional Structured Programming}

\author{Steven Obua}
\institute{}
\date{}


\maketitle

\begin{abstract}
The idea of functional programming has played a big role in shaping today's landscape of mainstream programming languages. 
Another concept that dominates the current programming style is Dijkstra's structured programming. Both concepts have been successfully married, for example in the programming language Scala. This paper proposes how the same can be achieved for structured programming and \emph{purely} functional programming via the notion of \emph{linear scope}. 
One advantage of this proposal is that mainstream programmers can reap the benefits of purely functional programming like easily exploitable parallelism  while using familiar structured programming syntax and without knowing concepts like monads. A second advantage is that professional purely functional programmers can often avoid hard to read functional code by using structured programming syntax that is often easier to parse mentally.
\end{abstract}

\section{Introduction}
\emph{Purely functional programming} (PFP) has a chance of becoming very popular for the simple reason that we now have laptops with four cores and more. The promise of PFP is that because there are no side-effects, no destructive updates, and no shared mutable state, partitioning a program into pieces that run in parallel becomes straightforward. 

Another consequence of the freedom from impure language constructs is that reasoning about program correctness, both formally and informally, becomes much easier in PFP languages than in, say, imperative languages. Therefore it is not surprising that PFP is popular within the theorem proving community.  For example, the source code of the interactive theorem proving assistant Isabelle~\cite{isabelle} is mostly written in a purely functional style. Outside of such specialty communities though, PFP clearly has not reached the mainstream yet. 

A programming paradigm that pervades today's mainstream is Dijkstra's \emph{structured programming}~\cite{structuredprogramming} (SP). Most young programmers even do not know the term structured programming anymore, but anyway still construct their object-oriented programs out of building blocks like \bsrc{if}-branches and \bsrc{while}-loops.

Interestingly, the PFP community largely rejects SP because it smells of side-effects, destructive updates, and mutable state, just the things a purely functional programmer wants to avoid. As an example, let us examine the Isabelle (version 2009-2) source code.
Discounting blank lines,  it consists of about 140000 lines of Standard ML~\cite{standardml} (SML) code. Yet, only ten of those lines use the \bsrc{while} keyword of SML! Furthermore, five out of those ten lines are part of Isabelle's system level code, and a further three lines stem from the author of this paper trying to circumvent missing tail-recursion optimization. The reason for this sparse use of \bsrc{while} is clear: in order to use \bsrc{while} in SML one must also use reference cells which are the embodiment of the small amount of impurity still left in SML. 

The easiest way to make PFP more mainstream might be to make SP, which already is part of the mainstream, an integral part of PFP! This is what this paper is about. Our central tool for such a unification of PFP and SP is the notion of \emph{linear scope}. Linear scope makes heavy use of \emph{shadowing}, therefore we first look at shadowing and its treatment in other languages that draw on functional programming, like Erlang and Scala. We then present the syntax of a toy language called  \emph{Mini Babel-17} to prepare a proper playground for the introduction of linear scope. First we concentrate on how linear scope interacts with the sequencing and nesting of statements. From there the extension to conditionals and loops is straightforward.
Finally we give a formal semantics for Mini Babel-17 and hence also for linear scope.

\section{Shadowing is Purely Functional}\label{sec:shadowing}
Here is how you could code in SML the function $x \mapsto x^4$:
\begin{babellisting}
fn x => let val x = x * x in x * x end
\end{babellisting}
There is no doubt that the above denotes a pure function. The fact that the introduction of the variable \bsrc{x} via \bsrc{val x = x * x} \emph{shadows} the previous binding of \bsrc{x} in \bsrc{fn x} might make it look a little bit more unusual than the more common 
\begin{babellisting}
fn x => let val y = x * x in y * y end,
\end{babellisting}
but of course both denotations are equivalent. Rewriting both functions in De Bruijn notation~\cite{debruijn} would actually yield the exact same closed term.

Yet it seems that the conception that shadowing is somehow wrong lies at the heart of why PFP and SP do not overlap in the mind of many programmers. 

An instance where shadowing is forbidden in order to obtain a notion of pure variables is the programming language Erlang which features \emph{single-assignment} of variables. Quoting the inventor of Erlang~\cite[p. 29]{erlang}: 
\begin{quote}
When Erlang sees a statement such as X = 1234, it binds the variable X to the value 1234. Before being bound, X could take any value: it's just an empty hole waiting to be filled. However, once it gets a value, it holds on to it forever.
\end{quote}
Clearly in Erlang shadowing is a victim of the idea that variables are not just names bound to values, but that \emph{the variables themselves are the state}.

Something similar can be observed in the programming language Scala~\cite{scala}. Scala combines functional and structured programming in an elegant fashion. But when it comes to integrate \emph{purely} functional programming and SP, Scala does not go all the way: it also forbids shadowing. For example, the following Scala function implements $x \mapsto x^8$,
\begin{babellisting}
(x : Int) => { val y = x*x; val z = y*y; z*z },
\end{babellisting}
but both of the following expressions are illegal in Scala:
\begin{babellisting}
(x : Int) => { val y = x*x; val y = y*y; y*y },
(x : Int) => { val y = x*x; y = y*y; y*y }.
\end{babellisting}
The last expression can be turned into a legal Scala expression by replacing the keyword \bsrc{val}, which introduces immutable variables, with the keyword \bsrc{var}, which introduces mutable variables:
\begin{babellisting}
(x : Int) => { var y = x*x; y = y*y; y*y }.
\end{babellisting}
It might seem that after all, shadowing in Scala is possible! But this is not the case. That \bsrc{var}
behaves differently than shadowing can easily be checked:
\begin{babellisting}
(x : Int) => {  var y = x*x
                val h = () => y
                y = y*y
                h() * y  }
\end{babellisting}
also implements $x \mapsto x^8$. With shadowing, we would expect above function to implement $x \mapsto x^6$.

\section{Syntax of Mini Babel-17}
\emph{Babel-17}~\cite{babel17} is a new dynamically-typed programming language in the making which is being developed by the author of this paper. One of its main features is that it combines purely functional programming and structured programming, building on the key observation that shadowing is purely functional. For illustration purposes we use a simplified version of a subset of Babel-17, which we call \emph{Mini Babel-17}, as a proposal of how a purely functional structured programming language could look like. An implementation of  Mini Babel-17 is available at~\cite{minibabel17}.

A \emph{block} in Mini Babel-17 is a sequence of \emph{statements}:
\begin{babellisting}
$\textsl{block}$ $\longrightarrow$ $\textsl{statement}_1$
              $\vdots$
          $\textsl{statement}_n$

\end{babellisting}
Several statements within a single line are separated via semicolons. There are seven kinds of statements: 
\begin{babellisting}
$\textsl{statement}$ $\longrightarrow$ $\textsl{val-statement}$
            $|$  $\textsl{assign-statement}$
            $|$  $\textsl{yield-statement}$
            $|$  $\textsl{if-statement}$
            $|$  $\textsl{while-statement}$
            $|$  $\textsl{for-statement}$
            $|$  $\textsl{block-statement}$            

$\textsl{val-statement}$ $\longrightarrow$ val $\textsl{identifier}$ = $\textsl{expression}$
$\textsl{val-assign-statement}$ $\longrightarrow$ $\textsl{identifier}$ = $\textsl{expression}$
$\textsl{yield-statement}$ $\longrightarrow$ yield $\textsl{expression}$ 
$\textsl{if-statement}$ $\longrightarrow$ if $\textsl{simple-expression}$ then $\textsl{block}$ else $\textsl{block}$ end 
$\textsl{while-statement}$ $\longrightarrow$ while $\textsl{simple-expression}$ do $\textsl{block}$ end 
$\textsl{for-statement}$ $\longrightarrow$ for $\textsl{identifier}$ in $\textsl{simple-expression}$ do $\textsl{block}$ end 
$\textsl{block-statement}$ $\longrightarrow$ begin $\textsl{block}$ end 
\end{babellisting}
If the last statement of a \emph{block} is  a \emph{yield-statement}, then the \bsrc{yield} keyword may be dropped in that statement. 

A \emph{simple-expression} is an expression like it can be found in most other functional languages, i.e. it can be an integer, a boolean, an identifier, an anonymous function, function application, or some operation on  \emph{expressions} like function application, multiplication or comparison:
\begin{babellisting}
$\textsl{simple-expression}$ $\longrightarrow$ 
     $\textsl{integer}$ $|$ $\textsl{boolean}$ $|$ $\textsl{identifier}$ 
    $|$ $\textsl{identifier}$ => $\textsl{expression}$
    $|$ $\textsl{expression}_1$ $\textsl{expression}_2$    
    $|$ $\textsl{expression}_1$ * $\textsl{expression}_2$
    $|$ $\textsl{expression}_1$ == $\textsl{expression}_2$    
    $\vdots$
\end{babellisting}
An \emph{expression} is either a \emph{simple-expression} or a statement:
\begin{babellisting}
$\textsl{expression}$ $\longrightarrow$ $\textsl{simple-expression}$ 
             $|$  $\textsl{if-statement}$
             $|$  $\textsl{while-statement}$
             $|$  $\textsl{for-statement}$
             $|$  $\textsl{block-statement}$            
\end{babellisting}

\section{Linear Scope}
Let us gain a first intuitive understanding of Mini Babel-17 before formally introducing its semantics.
Here is how you could denote $x \mapsto x^8$ in Mini Babel-17:
\begin{babellisting}
x => begin val y = x*x; val z = y*y; z*z end
\end{babellisting}
This looks pretty much like the Scala denotation of  $x \mapsto x^8$ from Section~\ref{sec:shadowing}.
But because Mini Babel-17 is designed so that shadowing of variables is allowed, an equivalent notation is:
\begin{babellisting}
x => begin val x = x*x; val x = x*x; x*x end
\end{babellisting}
The central idea of Mini Babel-17 is the notion of \emph{linear scope}. Whenever an identifier x is in linear scope, it is allowed to rebind 
x to a new value, and \emph{that rebinding will affect all other \emph{later} lookups of $x$ that happen within its normal lexical scope}. The rebinding is done via a \emph{val-assign-statement}.

The linear scope of a variable is contained in the usual lexical scope of that variable.
The linear scope of a variable x starts 
\begin{itemize}
\item in the statement after the \emph{val-statement} that defines $x$, or 
\item in the first statement of an anonymous function that binds $x$ as a function argument, if the body of that function is a block, or
\item in the first statement of the block of a \emph{for-loop} where x is the identifier bound by that loop.
\end{itemize}
It continues throughout the rest of the block unless a new linear scope for $x$ starts. It does extend into nested blocks and statements, but not into \emph{simple-expressions}. The reason for this is that blocks and statements are ordered sequentially, but there is no natural order for the evaluation of the components of a \emph{simple-expression}. 

Using the linear scope rules of Mini Babel-17, the above function can also be encoded as
\begin{babellisting}
x => begin x = x*x; x = x*x; x*x end 
\end{babellisting}

If there are no nested blocks involved, then linear scope is no big deal. It is just a fancy way of saying that when in a \emph{val-statement} the variable being defined shadows a previously defined variable, often it is ok to drop the \emph{val} keyword, effectively turning a \emph{val-statement} into a \emph{val-assign-statement}. 

But with nested blocks, linear scope becomes important:
\begin{center}
\begin{tabular}{c|cc|cc}
\begin{babellisting}
val x = 2
begin
  val y = x*x
  val x = y
end
x+x  
\end{babellisting}
&
&
\begin{babellisting}
val x = 2
begin
  val y = x*x
  x = y
end
x+x 
\end{babellisting}
&
&
\begin{babellisting}
val x = 2
begin
  val y = x*x
  val x = 0
  x = y
end
x+x 
\end{babellisting}
\\\hline
evaluates to 4 & & evaluates to 8 & & evaluates to 4
\end{tabular}
\end{center}
The left and right programs both evaluate to 4 because the \bsrc{begin} ... \bsrc{end} block is superfluous as none of its statements have any effect in the outer scope.  The middle program evaluates to 8, though, because the rebinding \bsrc{x = y} effects all later lookups in the lexical scope of  that x that has been introduced via \bsrc{val x = 2}, and \bsrc{x+x} certainly is such a later lookup. 

Maybe the rules of linear scope sound confusing at first. But they really are not. Just replace in your mind in the above three programs the \bsrc{val}s by \bsrc{var}s and view them as imperative programs. What value would you assign now to each program?

Let us also recode the last Scala expression of Section~\ref{sec:shadowing} as a Mini Babel-17 expression:
\begin{babellisting}
x => begin
        val y = x*x
        val h = dummy => y
        y = y*y
        h 0 * y
      end
\end{babellisting} 
Mini Babel-17 is purely functional, therefore the value of h is of course not changed by the rebinding \bsrc{y = y*y} which affects only \emph{later} lookups of y. Thus the above expression implements $x \mapsto x^6$, not $x \mapsto x^8$.

\section{Conditionals and Loops}
Conditionals and especially loops are the meat of structured programming. With linear scope, they are easily seen also as part of purely functional programming. All we need to do is to apply linear scoping rules to the nested blocks that the \emph{if}-, \emph{while-} and \emph{for-statements} consist of. For example, this is how you can encode the subtraction based Euclidean algorithm for two non-negative integers in Mini Babel-17:
\begin{babellisting}
a => b =>
    if a == 0 then
      b
    else 
      val a = a
      while b != 0 do
        if a > b then
          a = a - b
        else
          b = b - a
        end
      end
      a
    end
\end{babellisting}
Note the line \bsrc{val a = a} which on first sight seems to be superfluous. But while the linear scope of \bsrc{b} encompasses the whole
function body, the linear scope of \bsrc{a} does not, because linear scope does not extend into \emph{simple-expressions}.
If Mini Babel-17 had pattern matching, the line \bsrc{val a = a} could be avoided by starting the function definition with 
\begin{babellisting}
[a,  b] =>
    $\vdots$
\end{babellisting}
instead.

\section{Semantics of Mini Babel-17}
In this section we define an operational semantics for Mini Babel-17 by building a Mini Babel-17 interpreter written in Standard ML\footnote{
The original SML sources of the interpreter and all Mini Babel-17 programs of this paper are available online~\cite{babel17}.}. 

First we represent the grammar of Mini Babel-17 as SML datatypes:
\begin{babellisting}
datatype block = Block of statement list
and statement = 
       SVal of identifier * expression
     | SAssign of identifier * expression
     | SYield of expression
     | SIf of simple_expression * block * block
     | SWhile of simple_expression * block
     | SFor of identifier * simple_expression * block
     | SBlock of block
and expression =
       ESimple of simple_expression
     | EBlock of statement
and simple_expression =
       EInt of int  | EBool of bool | EId of identifier
     | EFun of identifier * expression
     | EBinOp of (value * value -> value) * 
                  expression * expression
and identifier = Id of string
\end{babellisting}
Note that function application, multiplication, comparison and so on are all described via the \bsrc{EBinOp} constructor by providing a suitable parameter of type \bsrc{value * value -> value}. The type \bsrc{value} represents all values that can be the result of evaluating a Mini Babel-17 program:
\begin{babellisting}
datatype value = VBool of bool | VInt of int
               | VFun of value -> value 
               | VList of value list
\end{babellisting}

Mini Babel-17 wants to be both purely functional and structured; the most important ingredients of a purely functional program are expressions; the most important ingredients of an SP program are blocks and statements. This dilemma is resolved by treating statements as special expressions.

The interpreter defines the following evaluation functions:
\begin{babellisting}
eval_b : environment -> block -> environment * value list
eval_st : environment -> statement -> environment * value list
eval_e : environment -> expression -> value
eval_se : environment -> simple_expression -> value
\end{babellisting}
The evaluation of blocks and statements yields lists of values instead of single values, the block
\begin{babellisting}
begin yield 1; yield 2; 3 end
\end{babellisting}
for example evaluates to \bsrc{[1, 2, 3]}.

Consider the following Mini Babel-17 program:
\begin{babellisting}
val x = 0
begin x = 1; x end * begin val x = x + 2; x end
\end{babellisting}
It does not obey the linear scoping rules of Mini Babel-17 because x is not in linear scope in the \emph{val-assign-statement} \bsrc{x = 1}.
In such a situation, the exception Illformed is raised during evaluation. Furthermore, an exception TypeError is raised when for example the condition of an if-statement evaluates to a list instead of a boolean. Note by the way that the program
\begin{babellisting}
val x = 0
begin val x = 1; x end * begin val x = x + 2; x end
\end{babellisting}
is perfectly fine and evaluates to 2.

What does the environment look like? It is actually split into two parts, one part for those identifiers that have linear scope, and one part for identifiers that don't. The nonlinear part is a mapping from identifiers to values, the linear part a mapping from identifiers to reference cells of values. Both parts can be described by the polymorphic type 'a idmap:
\begin{babellisting}
type idmap = (string * 'a) list
fun lookup [] _ = raise Illformed
  | lookup ((t, x)::r) (Id s) = 
      if t = s then x else lookup r (Id s)
fun remove [] _ = []
  | remove ((t,x)::r) (Id s) = 
      if t = s then r else remove r (Id s)
fun insert m ((Id s),x) = (s,x)::(remove m (Id s))
\end{babellisting}
The type of environments is then introduced as follows:
\begin{babellisting}
type environment = value idmap * (value ref) idmap
fun deref [] = [] | deref ((s, vr)::m) = ((s,!vr)::(deref m))
fun freeze (nonlinear, linear) = (nonlinear@(deref linear), [])
fun bind (nonlinear, linear) (id,value) =
  (remove nonlinear id, insert linear (id, ref value))
fun rebind (env as (_, linear)) (id, value) =
  (lookup linear id := value; env)
\end{babellisting}
Note that bind returns a new environment, and rebind returns the same environment with a mutated linear part. The function freeze
turns all mutable linear bindings into immutable nonlinear ones.

Now we can give the definition of all evaluation functions:
\begin{babellisting}
fun eval_b env (Block []) = (env, [])
  | eval_b env (Block (s::r)) =
    let
      val (env', values_s) = eval_st env s
      val (env'', values_r) = eval_b env' (Block r)
    in (env'', values_s @ values_r) end
and eval_nestedb env b =
    let
      val (_, values) = eval_b env b
    in (env, values) end
and eval_st env (SVal (id, e)) =
    let
      val value = eval_e env e
    in (bind env (id, value), []) end
  | eval_st env (SAssign (id, e)) =
    let
      val value = eval_e env e
    in (rebind env (id, value), []) end
  | eval_st env (SYield e) =
    let
      val value = eval_e env e
    in (env, [value]) end
  | eval_st env (SBlock b) = eval_nestedb env b
  | eval_st env (SIf (cond, yes, no)) =
      (case eval_se env cond of
        VBool true => eval_nestedb env yes
      | VBool false => eval_nestedb env no
      | _ => raise TypeError)
  | eval_st env (loop as SWhile (cond, body)) =
      (case eval_se env cond of
        VBool true =>
          let
            val (_, values_1) = eval_b env body
            val (_, values_2) = eval_st env loop
          in (env, values_1 @ values_2) end
      | VBool false =>
          (env, [])
      | _ => raise TypeError)
  | eval_st env (SFor (id, list, body)) =
      (case eval_se env list of
         VList L => eval_for env id body L
       | _ => raise TypeError)
and eval_for env id body [] = (env, [])
  | eval_for env id body (x::xs) =
    let
      val (_, values_1) = eval_b (bind env (id,x)) body
      val (_, values_2) = eval_for env id body xs
    in (env, values_1@values_2) end
and eval_e env (ESimple se) = eval_se env se
  | eval_e env (EBlock s) =
      (case eval_b env (Block [s]) of
         (_, [a]) => a
       | (_, L) => VList L)
and eval_se env se = eval_simple (freeze env) se
and eval_simple env (EInt i) = VInt i
  | eval_simple env (EBool b) = VBool b
  | eval_simple env (EBinOp (f, a, b)) =
      f (eval_e env a, eval_e env b)
  | eval_simple (nonlinear, _) (EId id) = 
      lookup nonlinear id
  | eval_simple env (EFun (id, body)) =
      VFun (fn value =>
              eval_e (bind env (id, value)) body)
\end{babellisting}
Here is the evaluation function that computes the meaning of a Mini Babel-17 program, i.e. of a block:
\begin{babellisting}
eval : block -> value
fun eval prog = snd (eval_e ([], []) (EBlock (SBlock prog)))
\end{babellisting}

It is straightforward how to extract from above evaluation functions a wellformedness-criterion such that  if a Mini Babel-17 program is statically checked to be wellformed according to that criterion, no Illformed exception will be raised during the evaluation of the program:
\begin{babellisting}
val VALUE = VInt 0
fun check_b env (Block []) = env
  | check_b env (Block (s::r)) = 
    	check_b (check_st env s) (Block r)    
and check_st env (SVal (id, e)) =
      (check_e env e; bind env (id, VALUE))
  | check_st env (SAssign (id, e)) =
  	  (check_e env e; rebind env (id, VALUE))
  | check_st env (SYield e) = (check_e env e; env)
  | check_st env (SBlock b) = (check_b env b; env)
  | check_st env (SIf (cond, yes, no)) =
      (check_se env cond; 
       check_b env yes; check_b env no; env)
  | check_st env (loop as SWhile (cond, body)) =
      (check_se env cond; check_b env body; env)
  | check_st env (SFor (id, list, body)) =
      (check_se env list; 
       check_b (bind env (id, VALUE)) body; env)
and check_e env (ESimple se) = check_se env se
  | check_e env (EBlock s) = 
  	  (check_b env (Block [s]); ()) 
and check_se env se = check_simple (freeze env) se
and check_simple env (EInt i) = ()
  | check_simple env (EBool b) = ()
  | check_simple env (EBinOp (f, a, b)) =
      (check_e env a; check_e env b)
  | check_simple (nonlinear, _) (EId id) = 
      (lookup nonlinear id; ())
  | check_simple env (EFun (id, body)) =
  	  check_e (bind env (id, VALUE)) body	  
fun check prog = check_e ([], []) (EBlock (SBlock prog))
\end{babellisting}
The function \textsl{check} terminates because it is basically defined via primitive recursion on the structure of the program. Furthermore, the set of calls to \textsl{lookup} generated during an execution of \textsl{check prog} is clearly a superset of the set of calls to \textsl{lookup} generated during the execution of \textsl{eval prog}.
Therefore, if \textsl{check prog} does not raise an exception \textsl{Illformed}, then neither will \textsl{eval prog}.

\section{Loops or Functionals?}
With Mini Babel-17, you can freely choose between a programming style that uses loops and a programming style that puts its emphasis on the use of higher-order functionals. If you have an imperative background, you might start out with using loops everywhere, and then migrate slowly to the use of functionals like \emph{map} or \emph{fold} as your understanding of functional programming increases.

But even after your functional programming skills have matured, you might still choose to use loops in appropriate situations. Let us for example look at a function that takes a list of integers $[a_0, \ldots, a_n]$ and an integer $x$ as arguments and returns the list
\begin{displaymath}
	[q_0, \ldots, q_n] \quad \text{where} \quad q_k = \sum_{i=0}^k a_i\, x^i
\end{displaymath}
The implementation in Mini Babel-17 via a loop is straightforward, efficient and even elegant:
\begin{babellisting}
m => x => begin
              val y = 0
              val p = 1
  	      for a in m do 
  	        y = y + a*p
	        p = p * x
  	        yield y
	      end
	   end
\end{babellisting}

\section{Related Work}
We have already mentioned how Scala also combines structured programming with functional programming, but fails to deliver a combination of structured programming and \emph{purely} functional programming. Actually, it should be possible to conservatively extend Scala so that linear scope for variables defined via \bsrc{val} is supported.

The work done on monads in the purely functional programming language Haskell~\cite{monads} has a superficial similarity with the work done in this paper. With monads it is possible to formulate sequences of (possibly shadowing) assignments, and with the help of monad transformers even loops can be modeled. But in order to understand and effectively use monads a solid background in functional programming is useful, if not even required; linear scope on the other hand is understood intuitively by programmers with a mostly imperative background, because Mini Babel-17 programs can look just like imperative programs and do not introduce additional clutter like the need for lifting.
Actually, in Haskell monads are used to limit the influence of mutable state to a confined region of the code that can be recognized by its type; the work in this paper has  the entirely different focus of trying to merge the structured and purely functional programming style as seamlessly as possible.

This work is not directly connected to work done on linear or uniqueness types~\cite{lineartypes}. Of course one might think about applying uniqueness typing to Mini Babel-17, but Mini Babel-17 itself is dynamically-typed and its values are persistent and can be passed around without any restrictions.

\section{Conclusion}
The current separation between SP and PFP is an artificial one. There is no good reason anymore why SP should not be used where appropiate for the sequential parts of a purely functional program except the personal preference of the programmer.  The purpose of Mini Babel-17 is to show the importance of linear scope for unifying SP and PFP. Babel-17 incorporates further important features for purely functional and structured programming like mutually recursive functions, pattern matching, exceptions, objects, memoization, concurrency, laziness, and more syntactic sugar. 


\bibliographystyle{abbrvnat}

\begin{thebibliography}{10}

\bibitem{erlang}Armstrong. \newblock\emph{Programming Erlang: Software for a Concurrent World}. \newblock Pragmatic Bookshelf, 2007.

\bibitem{structuredprogramming}Dahl, Dijkstra, Hoare. \newblock\emph{Structured Programming.} 
\newblock Academic Press, London, 1972.

\bibitem{debruijn}De Bruijn, Govert. \newblock\emph{A survey of the project AUTOMATH}. \newblock  In Hindley and Seldin (editors), \emph{To H. B. Curry: Essays on Combinatory Logic, Lambda Calculus and Formalism}, Academic Press, 1980.

\bibitem{monads}Jones, Wadler. \newblock Imperative Functional Programming. \newblock\emph{ACM Symposium on Principles of Programming Languages}, 1993, pp 71-74. 

\bibitem{standardml}Milner; Tofte, Harper, MacQueen. \newblock\emph{The Definition of Standard ML.}\newblock MIT Press, 1997.

\bibitem{scala}Odersky, Spoon, Venners. \newblock\emph{Programming in Scala}.\newblock Artima Press, 2008.

\bibitem{lineartypes} Wadler. \newblock Linear types can change the world! \newblock\emph{In M. Broy and C. Jones, editors, Programming Concepts and Methods}. North Holland, Amsterdam, 1990.

\bibitem{babel17}Babel-17. \newblock \url{http://www.babel-17.com}

\bibitem{minibabel17}Mini Babel-17. \newblock \url{http://www.babel-17.com/Mini-Babel-17}

\bibitem{isabelle}Isabelle. \newblock \url{http://isabelle.in.tum.de}

\bibitem{babel17novel}Delany. \newblock\emph{Babel-17}. \newblock Ace Books, 1966.

\end{thebibliography}


\end{document}